\documentstyle[12pt]{article}
\author{V.~A.~Slobodenyuk \thanks{E-mail: slob@themp.univ.simbirsk.su} \\
{\it Physical-Technical Department} \\
{\it Ulyanovsk State University } \\
{\it L.Tostogo str. 42, 432700 Ulyanovsk, Russian Federation}}
\title{On convergence of the Schwinger --- DeWitt expansion}
\date{}

\begin{document}

\maketitle

\begin{abstract}
The Schwinger --- DeWitt expansion for the
evolution operator kernel of the Schr\"odinger equation
is studied for convergence. It is established that divergence
of this expansion which is usually implied for all
continuous potentials, excluding ones of the form
$V(q)=aq^2+bq+c$, really takes place only if the coupling
constant $g$ is treated as independent variable. But the
expansion may be convergent for some kinds of the potentials
and for some discrete values of the charge, if the latter is
considered as fixed parameter. Class of such potentials is
interesting because inside of it the property of discreteness
of the charge in the nature is reproduced in the theory in
natural way.
\end{abstract}

\newpage

\bigskip

\section{Introduction}

The short-time Schwinger --- DeWitt expansion is used in the
quantum theory for various purposes~\cite{Sch,DeW1,DeW2,BOG,OF}.
As usually it is treated as asymptotic one, so as other
expansions in different parameters: conventional perturbation
theory~\cite{BW,Lip}, the WKB-expansion, $1/n$-expansion~\cite{PSS}
etc. We mean under the Schwinger --- DeWitt expansion
following representation of the solution of the Schr\"odinger equation
for the evolution operator kernel
  \begin{equation} \label{f1.1}
  i \frac{\partial}{\partial t} \langle \vec q\,',t\mid \vec q,0 \rangle =
  -\frac{1}{2} \sum_{\nu=1}^3 \frac{\partial^2}{\partial q'^2_{\nu}}
  \langle \vec q\,',t\mid \vec q,0 \rangle + V(\vec q\,')
  \langle \vec q\,',t\mid \vec q,0 \rangle
  \end{equation}
with initial condition
  \begin{equation} \label{f1.2}
  \langle \vec q\,',t=0\mid \vec q,0 \rangle = \delta (\vec q\,'-\vec q).
  \end{equation}
The kernel $\langle \vec q\,',t\mid \vec q,0 \rangle$ is written as
  \begin{equation} \label{f1.3}
  \langle \vec q\,',t\mid \vec q,0 \rangle = \frac{1}{(2\pi it)^{3/2}}
  \exp \left\{i \frac{(\vec q\,'-\vec q)^2}{2t} \right\}
  F(t;\vec q\,',\vec q),
  \end{equation}
and $F$ according to~\cite{Sch,DeW1,DeW2,S1,S2} is expanded in
powers of $t$
  \begin{equation} \label{f1.4}
  F(t;\vec q\,',\vec q) = \sum_{n=0}^{\infty} (it)^n a_n(\vec q\,',\vec q),
  \end{equation}
Here and everywhere below dimensionless values defined in obvious
manner are used. The potential $V(\vec q)$ is continuous function.

Factor in front of $F$ in R.h.s. of~(\ref{f1.3}) is the kernel for
the free theory, i.e. for $V\equiv 0$. Behavior of relation
$ \langle \vec q\,',t\mid \vec q,0 \rangle /
  \langle \vec q\,',t\mid \vec q,0 \rangle \Big|_{V=0} \Big.$
when $t \to 0$ was studied in~\cite{Martin} for wide class of
potentials and for $t=-i \tau, \ \tau >0$. It was established that
this relation tends to 1 for $\tau \to 0$. This fact may serve as
justification of representation~(\ref{f1.4}) with $a_0=1$.

Because the expansion~(\ref{f1.4}) is usually considered as
asymptotic, it is naturally that only the problem of character
of asymptotic growth was studied. E.g., estimates from above
were obtained for the coefficients $a_n$~\cite{OF,S2}. These
estimates show that $a_n < \Gamma(bn)$ ($0<b\le 1$) as
$n \to \infty$. But in general case they do
not prove that divergence certainly takes place for every
potential. Specifically,
in~\cite{S3,TMF2} some potentials were established for which the
Schwinger --- DeWitt expansion is convergent for definite
values of the charge. So, it is interesting to study the
problem: when convergence is possible and when is not?
Namely this problem is under consideration at present paper.

\section{Quantum mechanics in one-dimensional space}

\subsection{General prescription}

Let us consider the Cauchy problem for the evolution kernel for
the Schr\"odinger equation in one-dimensional space
  \begin{equation} \label{f2.1}
  i\frac{\partial}{\partial t} \langle q',t\mid q,0 \rangle =
  -\frac{1}{2} \frac{\partial^2}{\partial q'^2}
  \langle q',t\mid q,0 \rangle + V(q') \langle q',t\mid q,0 \rangle,
  \end{equation}
  \begin{equation} \label{f2.2}
  \langle q',t=0\mid q,0 \rangle = \delta (q'-q).
  \end{equation}
We represent the kernel in the form
  \begin{equation} \label{f2.3}
  \langle q',t\mid q,0 \rangle = \frac{1}{\sqrt{2\pi it}}
  \exp \left\{i \frac{(q'-q)^2}{2t} \right\} F(t;q',q),
  \end{equation}
where $F$ is given by the short-time expansion
  \begin{equation} \label{f2.4}
  F(t;q',q) = \sum_{n=0}^{\infty} (it)^n a_n(q',q).
  \end{equation}
The coefficient functions $a_n(q',q)$ may be determined from
the sequence of recurrent relations
  \begin{equation} \label{f2.5}
  a_0(q',q)=1,
  \end{equation}
  \begin{equation} \label{f2.6}
  na_1(q',q) + (q'-q) \frac{\partial a_1(q',q)}{\partial q'} =
                                       a_1(q',q')= -V(q'),
  \end{equation}
and
  \begin{equation} \label{f2.7}
  na_n(q',q) + (q'-q) \frac{\partial a_n(q',q)}{\partial q'}
  = \frac{1}{2} \frac{\partial ^2 a_{n-1}(q',q)}{\partial q'^2} -
  V(q')a_{n-1}(q',q)
  \end{equation}
for $n>1$.
Eqs.~(\ref{f2.5})--(\ref{f2.7}) show that $a_n$ can be calculated
via $a_{n-1}$ by means of integral relations
  \begin{equation} \label{f2.8}
  a_n(q',q)= \int \limits_0^1 \eta^{n-1} d\eta
  \left\{ \frac{1}{2} \frac{\partial ^2}{\partial x^2} -
  V(x) \right\} a_{n-1}(x,q)\Biggl.
  \Biggr|_{x=q+(q'-q)\eta}.
  \end{equation}
Combinations of eqs.~(\ref{f2.8}) for different numbers
$n$ allows us to represent $a_n$ for given $n$ through the
potential $V$ and its derivatives
  \begin{eqnarray} \label{f2.9}
  a_n(q',q)&=& -\int\limits_0^1 \eta_n^{n-1} d\eta_n
  \int\limits_0^1 \eta_{n-1}^{n-2} d\eta_{n-1} \dots
  \int\limits_0^1 \eta_2 d\eta_2 \int\limits_0^1 d\eta_1
  \left\{ \frac{1}{2} \frac{\partial^2}{\partial x_n^2} -V(x_n)\right\}
  \times \nonumber \\ &&
  \left\{ \frac{1}{2} \frac{\partial^2}{\partial x_{n-1}^2}
                                   -V(x_{n-1})\right\} \dots
  \left\{\frac{1}{2} \frac{\partial^2}{\partial x_2^2}  -V(x_2)\right\}
  V(x_1).
  \end{eqnarray}
Here $x_i=q+(x_{i+1}-q)\eta_i$, $x_{n+1}=q'$. Derivatives with
respect to different $x_i$ may be easily connected with each
other
  \begin{equation} \label{f2.10}
  \frac{\partial}{\partial x_i}=\eta_{i-1} \frac{\partial}{\partial x_{i-1}}
  =\eta_{i-1} \eta_{i-2} \frac{\partial}{\partial x_{i-2}}
  \end{equation}
etc. We use~(\ref{f2.9}) for evaluation the behavior of $a_n$
for $n \to \infty$.

The potential $V(q)$ contains the coupling constant $g$ as
multiplier. If $g$ is independent variable, as it is treated
in conventional perturbation theory, then for convergence of
the expansion~(\ref{f2.4}) one is to demand that contribution
into $a_n$ proportional to $g^l$ behaves itself for every $l$
such that the series in powers of $t$ is convergent. I.e.,
cancellations between contributions into $a_n$, containing
$g$ in different powers are not possible. So, it is convenient
to calculate not hole coefficients $a_n$, but only the terms
containing $g^l$, which we denote as $a_n^l$.

The representation~(\ref{f2.9}) allows us to analyse the
structure of $a_n^l$ and to understand sources of arising the
factorial growth of $a_n^l$ for $n \to \infty$. We start from
previous consideration.

\subsection{Previous consideration}

If one opens brackets in~(\ref{f2.9}) and takes only terms
proportional to $g^l$, i.e., containing $l$ multipliers of
type $V(x_i)$, then one gets
  $$\frac{(n-1)!}{(n-l)!(l-1)!}$$
addends of the form
  \begin{equation} \label{f2.11}
  \int \limits_0^1 {\cal D}_n \eta \; \frac{1}{2^{n-l}}
  W(x_n)W(x_{n-1}) \dots W(x_2)V(x_1),
  \end{equation}
where $W(x_i)$ is either $V(x_i)$ or
               ${\displaystyle \frac{\partial^2}{\partial x_i^2}}$,
  $${\cal D}_n \eta \equiv \eta_n^{n-1} d\eta_n
  \eta_{n-1}^{n-2} d\eta_{n-1} \dots \eta_2 d\eta_2 d\eta_1 $$
And besides number of multipliers $W(x_i)=V(x_i)$ at every
term is equal to $l-1$, and number of multipliers
${\displaystyle W(x_i)=\frac{\partial^2}{\partial x_i^2}}$
is equal to $n-l$.
Total order of derivatives acting on all $l$ multipliers
$V(x_i)$ is equal to $2(n-l)$.

To produce differentiating in~(\ref{f2.11}) it is convenient to
represent the operators
${\displaystyle \frac{\partial}{\partial x_i}}$ via other
operators ${\displaystyle \frac{\partial}{\partial x_j}}$ with a help of
relations of type~(\ref{f2.10}) so that only derivatives
  $$\frac{d^m V(x_j)}{dx_j^m} \equiv V^{(m)}(x_j)$$
will be present at all expressions.
During this transform additional
factors of type $\eta_i^{k_i}$ will appear (we call
them $\eta$-factors). From every term of the form~(\ref{f2.11})
one will get $l^{2(n-l)}$ terms of type
  \begin{equation} \label{f2.12}
  \frac{(-1)^l}{2^{n-l}}\int \limits_0^1 {\cal D}_n \eta\;
  \eta_{n-1}^{k_{n-1}} \dots \eta_2^{k_2} \eta_1^{k_1}
  V^{(m_1)}(x_{i_1}) \dots V^{(m_l)}(x_{i_l}).
  \end{equation}
Let us evaluate the contribution of $\eta$-factors
$\eta_{n-1}^{k_{n-1}} \dots \eta_2^{k_2} \eta_1^{k_1}$
into~(\ref{f2.12}). We put for a moment all $V^{(m)}(x)$
independent on $\eta_i$ ($i=1, \dots, n$). Then the integrals
with respect to $\eta_i$ may give maximal factor $1/n!$ (when
in~(\ref{f2.12}) every $k_i=0$) and minimal factor
$(n-1)!/(2n-1)!$ (when every $k_i$ is equal to maximal possible
value $k_i=2(n-i)$). Because
  $$\frac{(n-1)!}{(2n-1)!} \sim \frac{1}{4^n n!}$$
as $n \to \infty$, then it is not essentially what exact
set of $k_i$ is used in~(\ref{f2.12}).

Among all terms of kind~(\ref{f2.11}) the following term
gives the largest contribution as $n \to \infty$
  \begin{equation} \label{f2.13}
  \frac{(-1)^l}{2^{n-l}}\int \limits_0^1 {\cal D}_n \eta\;
  \frac{\partial^{2(n-l)}}{\partial x_n^2 \dots \partial x_{l+1}^2}
  V(x_l) \dots V(x_1).
  \end{equation}
It is so, because after differentiating in~(\ref{f2.13}) it will
appear maximum number of the terms of type~(\ref{f2.12}). If
one ignores presence of $\eta$-factors evaluating~(\ref{f2.13}),
i.e., puts every $\eta_i$ equal to 1 (but not in
${\cal D}_n \eta$!), then one gets
  \begin{equation} \label{f2.14}
  \frac{(-1)^l}{2^{n-l}}\int \limits_0^1 {\cal D}_n \eta\;
  \sum_{\{ m_i \}} \frac{(2(n-l))!}{m_1! \dots m_l!}
  V^{(m_1)}(x_1) \dots V^{(m_l)}(x_l),
  \end{equation}
where following notation is used
  $$\sum_{\{ m_i \}} \equiv \sum_{m_1=0}^{2(n-l)}
  \sum_{m_2=0}^{2(n-l)-m_1} \dots
  \sum_{m_{l-1}=0}^{2(n-l)-m_1-\dots -m_{l-2}}
  \Biggl. \Biggr|_{m_l=2(n-l)-\sum\limits_{i=1}^{l-1} m_i}.$$

Suppose that in some domain the expansion
  \begin{equation} \label{f2.15}
  V(x)= \sum_{m=0}^{\infty} \frac{V^{(m)}(q)}{m!} (x-q)^m =
        \sum_{m=0}^{\infty} C_m(q) (x-q)^m
  \end{equation}
takes place. Then instead of~(\ref{f2.14}) we have
  \begin{equation} \label{f2.16}
  (-1)^l\frac{(2(n-l))!}{2^{n-l}} \int \limits_0^1 {\cal D}_n \eta\;
  \sum_{\{ m_i \}} C_{m_1}(x_1) \dots C_{m_l}(x_l).
  \end{equation}
Number of addends in $\sum_{\{ m_i \}}$ is equal to
  $$\frac{(2n-l-1)!}{(2(n-l))!(l-1)!} < 2^{2n-l-1}.$$
Hence, this cause cannot lead to asymptotic growth of type $n!$.

If we choose among all terms in~(\ref{f2.16}) only one with
$m_i=[2(n-l)/l], \; i=1, \dots, l$
([\dots] denotes integer part of number;
for the sake of brevity we will consider
that all $m_i$ are equal to each other; more strictly, it is
necessary to put some $m_i$ equal to $[2(n-l)/l]+1$,
because condition $\sum_{i=1}^l m_i=2(n-l)$ is to be
fulfilled, but for our asymptotic estimates this is not
essential) and evaluate the functions $C_m(x)$ by the
constants $C_m \equiv \max_{\{x\}} C_m(x)$, then we get for
this term estimate
  \begin{equation} \label{f2.17}
  \frac{(2(n-l))!}{2^{n-l}n!} C_{[2(n-l)/l]}^l.
  \end{equation}

Consider the contributions, containing $g$ in power
$l=[n \alpha]$, where $\alpha$ is any number from the
interval $0<\alpha<1/2$. For such $l$ the
contribution~(\ref{f2.17}) behaves itself for $n \to \infty$
as $\Gamma(n(1-2\alpha))$, which means that the
series~(\ref{f2.4}) diverges. So, we see that there are
contributions into $a_n$, which has factorial growth as
$n \to \infty$.

Now we accurately will study when such divergence really take
place. For proving of divergence of the expansion~(\ref{f2.4})
it is enough to prove that $a_n(q',q)$ rises as $\Gamma(bn)$
at least for any values of $q', \, q$ from analytic domain
of the function $V(q)$. Practically it is convenient to
consider $q'=q$. In this case in~(\ref{f2.16}) all $x_i=q$ and
$\sum_{\{ m_i \}} C_{m_1}(q) \dots C_{m_l}(q)$ does not
depend on integration variables $\eta_i$ and integrals can
be exactly calculated.

\subsection{Possible cancellations}

Generally speaking, different contributions into $a_n^l$,
having asymptotic growth of type $\Gamma(bn)$, may cancel
each other because of different signs of $V^{(m)}(q)$ for
different $m$, so that hole coefficient $a_n^l$ will not
have such behavior. At first, we show that such cancellations
really do not take place.

The most simple case is one when all $V^{(m)}(q)$ either have
the same sings or $V^{(m)}(q)=(-1)^m |V^{(m)}(q)|$. Then all
structures $V^{(m_1)}(q) \dots V^{(m_l)}(q)$ for every set of
$m_i$ have the same sign and no any cancellations can occur.

Consider general case, when the signs of $V^{(m)}(q)$ depend
on $m$ in arbitrary way. Evaluate $a_n^l(q,q)$
  \begin{equation} \label{f2.18}
  |a_n^l(q,q)|= \left| \sum_{\{ m_i \}} A^{nl}_{\{m_i\}}
  V^{(m_1)}(q) \dots V^{(m_l)}(q) \right|,
  \end{equation}
where the coefficients $A^{nl}_{\{m_i\}}$ take into
account contributions of all terms of kind~(\ref{f2.12})
(notation $\{m_i\}$ in index of $A$ is used instead of
$m_1, \dots, m_l$).
As it is clear from previous estimates, there are such
$A^{nl}_{\{m_i\}}$ that behave themselves as $\Gamma(n(1-2\alpha))$
when $n \to \infty$ for $l=[n\alpha], \; 0<\alpha<1/2$.
Assume that for some $q$, e.g., for $q=0$, the sings of derivatives
$V^{(m)}$ are such that $|a_n^l(0,0)|$ increase not strongly
then $n^c B^n$ ($c$ and $B$ are any positive constants) or even decrease
for $n \to \infty$. Let us test conservation of this property
for $q \neq 0$. We represent
  \begin{equation} \label{f2.19}
  V^{(m)}(q)= \sum_{k=0}^{\infty} \frac{V^{(m+k)}(0)}{k!} q^k
  \end{equation}
and substitute~(\ref{f2.19}) into~(\ref{f2.18})
  \begin{eqnarray} \label{f2.20}
  &&|a_n^l(q,q)|= \left| \sum_{k=0}^{\infty} q^k \sum_{\{ m_i \}}
  A^{nl}_{\{m_i\}} \times
  \right. \nonumber \\ && \left.
  \sum_{k_1=0}^{k} \sum_{k_2=0}^{k-k_1} \dots
  \sum_{k_{l-1}=0}^{k-k_1-\dots-k_{l-2}}
  \frac{V^{(m_1+k_1)}(0)}{k_1!} \dots \frac{V^{(m_l+k_l)}(0)}{k_l!}
  \right|_{k_l=k-\sum\limits_{i=1}^{l-1} k_i}.
  \end{eqnarray}
Because $q$ is independent variable then the coefficients
in front of every power of $q$ should have property mentioned
above.

The coefficient in front of $q^0$ is equal to
  \begin{equation} \label{f2.21}
  \sum_{\{ m_i \}} A^{nl}_{\{m_i\}} V^{(m_1)}(0) \dots V^{(m_l)}(0)
  \end{equation}
and according to our supposition it may increase not strongly
then $n^c B^n$ because of cancellations (nevertheless, among
$A^{nl}_{\{m_i\}}$ are such ones which increase as $\Gamma(n-2l)$).

Consider, for example, the coefficient in front of first power of $q$
  \begin{eqnarray} \label{f2.22}
  &&\sum_{\{ m_i \}} A^{nl}_{\{m_i\}} \left\{
  V^{(m_1+1)}(0)V^{(m_2)}(0) \dots V^{(m_l)}(0) +
  \right. \nonumber \\ && \left.
  V^{(m_1)}(0)V^{(m_2+1)}(0) \dots V^{(m_l)}(0) + \dots +
  V^{(m_1)}(0) \dots V^{(m_l+1)}(0) \right\} =
  \nonumber \\ &&
  l \sum_{\{ m_i \}} A^{nl}_{\{m_i\}}
  V^{(m_1+1)}(0)V^{(m_2)}(0) \dots V^{(m_l)}(0).
  \end{eqnarray}
In last equality the symmetry of $A^{nl}_{\{m_i\}}$ in indexes
$m_i$ for $q'=q$ is taken into account. Expression~(\ref{f2.22})
differs from~(\ref{f2.21}) (besides nonessential factor $l$)
only by replacement of $V^{(m_1)}(0)$ by $V^{(m_1+1)}(0)$. The
derivatives of order $m_1+1$ have for some $m_1$ the same sign
with $V^{(m_1)}(0)$, and for other $m_1$ --- opposite sing.
I.e., in~(\ref{f2.21}), where because of exact choice of signs
and values of derivatives $V^{(m_i)}(0)$ contributions rising
as factorial exactly cancel each other, the signs of many
terms have been arbitrary changed. So, initial exact tuning will
disappear and there will not be full compensation of rising
contributions. Even if in any special case sufficient
compensation of rising contributions will be accidentally
conserved for the terms considered, then we should have in
mind infinite number of terms with $k > 1$ in~(\ref{f2.20}).
There exist some terms among them for which disbalance
between contributions with opposite signs occur.
Hence, $|a_n^l(q,q)|$ will increase for
$q \neq 0$ as $\Gamma(n(1-2\alpha))$ even in case, when
$|a_n^l(0,0)|$ increases not strongly then $n^c B^n$.

\subsection{Proof of divergence}

Because there are no essential cancellations between different
contributions into $a_n^l(q,q)$ it is enough for proving
of divergence of the series~(\ref{f2.4}) to indicate any
term in $a_n^l(q,q)$ rising as factorial. The way of doing this
depends on analytical properties of the potential $V(q)$. At
first we consider the case when $V(q)$ is entire function of the
complex variable $q$. Then the series~(\ref{f2.15}) is convergent
for any $|q|<\infty$ and convergence range is infinite.

Let us take~(\ref{f2.13}) and, calculating expression of
type~(\ref{f2.14}), use $\eta$-factor in the form
$\eta_{n-1}^{2} \eta_{n-2}^{4} \dots \eta_2^{2(n-2)} \eta_1^{2(n-1)}$,
i.e., we change exact powers $k_i$ of the variables $\eta_i$
on maximal possible values $k_i=2(n-i)$, which gives estimate
from below for the contribution considered. We get
  \begin{eqnarray} \label{f2.23}
  &&\frac{1}{2^{n-l}}\int \limits_0^1 {\cal D}_n \eta\;
  \eta_{n-1}^{2} \eta_{n-2}^{4} \dots \eta_2^{2(n-2)} \eta_1^{2(n-1)}
  \sum_{\{m_i\}} \frac{(2(n-l))!}{m_1! \dots m_l!}
  V^{(m_1)}(q) \dots V^{(m_l)}(q) =
  \nonumber \\ &&
  \frac{1}{2^{n-l}} \frac{(n-1)!(2(n-l))!}{(2n-1)!}
   \sum_{\{m_i\}} C_{m_1}(q) \dots C_{m_l}(q).
  \end{eqnarray}
At the sum $\sum_{\{m_i\}}$ we take the term with
$m_i=[2(n-l)/l], \; i=1, \dots, l$ and consider the contribution
with $l=[n\alpha], \; 0<\alpha<1/2$. This gives for
$a_n^{[n\alpha]}$ asymptotic estimate
  \begin{equation} \label{f2.24}
  \frac{(n-1)![2n(1-\alpha)]!}{2^{n-[n\alpha]}(2n-1)!}
  C_{[2(1-\alpha)/\alpha]}^{[n\alpha]}(q) \sim \Gamma(n(1-2\alpha)),
  \end{equation}
which is fair for every entire function $V(q)$ so as for
function $V(q)$ analytical at any bounded domain of
variable $q$, because $C_{[2(1-\alpha)/\alpha]}(q)$ does
not depend on number $n$ and so behavior of Taylor's
coefficients $C_k(q)$ does not affect essentially on the
estimate~(\ref{f2.24}) (it can arise only the factor of type
$n^c B^n$, which does not change the qualitative character of
asymptotics~(\ref{f2.24})).

So, we established such contribution into $a_n$, that increases
as $\Gamma(n(1-2\alpha))$ and leads to divergence of the series~(\ref{f2.4}).
In realty there exist many of such contributions. One can choose
different $\alpha$, i.e., different $l$. The less $\alpha$, the more
strong asymptotic increase of corresponding contributions into
$a_n$ will take place. Nevertheless, one cannot put $\alpha=0$
because $l \ge 1$. To understand what is maximal growth of $a_n$
for $n \to \infty$ one is to admit $\alpha$ to decrease slowly
with increasing of $n$, e.g., as $\alpha=1/\log n$ (see~\cite{S2}). And
what's more, always $l=[n/\log n] > 1$ and $\Gamma (n(1-2 \log n))
\sim n!$ for $n \to \infty$.

We see that the coefficients of the expansion~(\ref{f2.4})
increase as $n!$
for $n \to \infty$ excluding cases when $V(q)$ is polynomial of
power $L$. In this case $V^{(m)}(q) \neq 0$ only for $m \le L$
and so contribution into $a_n^l$ is not equal to zero only if
$2(n-l) \le Ll$ or $L \ge 2n/(L+2)$. This gives additional
restriction on $\alpha$: $2/(L+2) \le \alpha < 1/2$. Hence,
maximal growth of $a_n^l$ is reached when $\alpha=2/(L+2)$ or
$l=[2n/(L+2)]$. It provides the behavior~\cite{S2}
  \begin{equation} \label{f2.25}
  |a_n| \sim \Gamma \left( n \frac{L-2}{L+2} \right).
  \end{equation}

For the potential $V(q)$, which is determined by the function
analytical at any bounded domain, corresponding considerations
are more simple. In this case the expansion~(\ref{f2.15})
converges at the circle of finite radius $R(q)$ and
$|C_m(q)| \sim 1/R^m(q)$ when $m \to \infty$.

We will calculate $a_n^1(q,q)$. It is clear from~(\ref{f2.8}), that
  \begin{eqnarray} \label{f2.26}
  &&a_n^1(q,q) = -\int \limits_0^1 {\cal D}_n \eta\; \frac{1}{2^{n-1}}
  \frac{\partial^{2(n-1)}}{\partial x_n^2 \dots \partial x_{2}^2}
  V(x_1) \Biggl. \Biggr|_{x_i=q} =
  \nonumber \\ &&
  -\frac{1}{2^{n-1}} \int \limits_0^1 {\cal D}_n \eta\;
  \eta_{n-1}^2 \eta_{n-2}^4 \dots \eta_1^{2(n-1)}
  V^{2(n-1)}(q)= - \frac{(n-1)!}{2^{n-1} (2n-1)} C_{2(n-1)}(q).
  \end{eqnarray}
Taking into account behavior of the coefficients $C_m(q)$,
we get estimate
  \begin{equation} \label{f2.27}
  |a_n^1(q,q)| \sim \frac{(n-1)!}{2^{n-1} (2n-1)}
  \frac{1}{R^{2(n-1)}(q)} \sim n!
  \end{equation}
for $n \to \infty$. Because $a_n^1$  contains only one term,
the problem of possible cancellations does not arise at all.

Thus, if one suppose that the charge $g$ is independent variable
and so cancellations between $a_n^l$ with different numbers $l$
do not occur, then for every potential $V(q)$ the coefficients
$a_n$ of the Schwinger --- DeWitt expansion~(\ref{f2.3}),
(\ref{f2.4}) behave themselves as $\Gamma(bn)$ when $n \to \infty$
and, hence, the expansion is always divergent (naturally,
excluding trivial case $V(q)=aq^2+bq+c$). The only way to get
convergent expansion is following. One is to consider the
charge $g$ not as independent variable, but as a fixed parameter.
Then cancellations between $a_n^l$ with different $l$ are
possible. For some potentials and for special values of the
charge $g$  these cancellations are sufficient to provide
convergence of the Schwinger --- DeWitt expansion. The
examples of such potentials was presented in~\cite{S3,TMF2}
  \begin{equation} \label{f2.28}
  V(q) = - \frac{g}{\cosh^2 q},
  \end{equation}
  \begin{equation} \label{f2.29}
  V(q) = \frac{g}{q^2},
  \end{equation}
  \begin{equation} \label{f2.30}
  V(q) = \frac{g}{\sinh^2 q},
  \end{equation}
  \begin{equation} \label{f2.31}
  V(q) = \frac{g}{\sin^2 q},
  \end{equation}
  \begin{equation} \label{f2.32}
  V(q) = aq^2 + \frac{g}{q^2}.
  \end{equation}

The potentials~(\ref{f2.29}) -- (\ref{f2.32}) have singularity
at $q=0$ which does not allow us to use for them directly the
formalism described above. But this formalism can be easily
modified for application to singular potentials~\cite{MPLA2}.
One should take instead of initial condition~(\ref{f2.2})
the following one
  \begin{equation} \label{f2.33}
  \langle q',t=0\mid q,0 \rangle = \delta (q'-q) + A \delta (q'+q)
  \end{equation}
which may provide fulfillment of boundary condition for
the wave function $\psi(q)$ at $q=0$ ($\psi(q)$ should vanish
at $q=0$) by appropriate choice of constant $A$. Constant
$A$ is determined by requirement that the kernel does not have
singularity at $q=0$ or $q'=0$ ($t \ne 0$). In correspondence
with~(\ref{f2.33}) the kernel is represented through two
functions $F^{(\pm)}$ as
  \begin{eqnarray} \label{f2.34}
  \langle q',t\mid q,0 \rangle &=& \frac{1}{\sqrt{2\pi it}}
  \exp \left\{i \frac{(q'-q)^2}{2t} \right\} F^{(-)}(t;q',q) +
  \nonumber \\ &&
  A \frac{1}{\sqrt{2\pi it}}
  \exp \left\{i \frac{(q'+q)^2}{2t} \right\} F^{(+)}(t;q',q),
  \end{eqnarray}
where $F^{(\pm)}$ can be expanded analogously to~(\ref{f2.4}).

The function $F^{(-)}$ may be calculated in the same way as
ordinary function $F$ discussed before. Calculation of $F^{(+)}$
slightly differs from one of $F^{(-)}$, but in our consideration
function $F^{(+)}$ is not essential. If we wish to prove
divergence of representation~(\ref{f2.34}), (\ref{f2.4}) for
the kernel it is enough to prove divergence of the
expansion~(\ref{f2.4}) only for function $F^{(-)}$.
Independently on the behavior of $F^{(+)}$
representation~(\ref{f2.34}), (\ref{f2.4}) will be divergent
in this case. So, statement about divergence of the
Schwinger --- DeWitt expansion for arbitrary charge $g$
remains fair and for singular potentials too.

\section{Quantum mechanics in three-dimensional space}

\subsection{General consideration}

Formalism used in the one-dimensional case can be easily transferred
on the three-dimensional space. Let us write corresponding
formulas. Equations for $a_n(\vec q\,',\vec q)$ defined
by~(\ref{f1.1})--(\ref{f1.4}) are
  \begin{equation} \label{f3.1}
  na_1(\vec q\,',\vec q) + \sum_{\nu=1}^3 (q'_{\nu}-q_{\nu})
         \frac{\partial a_1(\vec q\,',\vec q)}{\partial q'_{\nu}} =
                      a_1(\vec q\,',\vec q\,')= -V(\vec q\,'),
  \end{equation}
and
  \begin{equation} \label{f3.2}
  na_n(\vec q\,',\vec q) + \sum_{\nu=1}^3 (q'_{\nu}-q_{\nu})
        \frac{\partial a_n(\vec q\,',\vec q)}{\partial q'_{\nu}}
  = \frac{1}{2} \Delta_{\vec q\,'} a_{n-1}(\vec q\,',\vec q) -
  V(\vec q\,')a_{n-1}(\vec q\,',\vec q)
  \end{equation}
for $n>1$.
The solution of~(\ref{f3.1})--(\ref{f3.2}) has a form
  \begin{eqnarray} \label{f3.3}
  a_n(\vec q\,',\vec q)&=& -\int\limits_0^1 \eta_n^{n-1} d\eta_n
  \int\limits_0^1 \eta_{n-1}^{n-2} d\eta_{n-1} \dots
  \int\limits_0^1 \eta_2 d\eta_2 \int\limits_0^1 d\eta_1
  \left\{ \frac{1}{2} \Delta_n -V(\vec x_n)\right\}
  \times \nonumber \\ &&
  \left\{ \frac{1}{2} \Delta_{n-1} -V(\vec x_{n-1})\right\} \dots
  \left\{ \frac{1}{2} \Delta_2 -V(\vec x_2)\right\}  V(\vec x_1),
  \end{eqnarray}
where $\vec x_i=\vec q+(\vec x_{i+1}-\vec q)\eta_i$,
$\vec x_{n+1}=\vec q\,'$, ${\displaystyle \Delta_i = \sum_{\nu=1}^3
\frac{\partial^2}{\partial x^2_{i\nu}} }$ is Laplasian acting
on functions of the variable $\vec x_i$, index $\nu=1,2,3$ is
Cartesian index. Correspondence between various differential
operators is analogous to~(\ref{f2.10}).

Because structure of the representation~(\ref{f3.3}) is the
same as one for~(\ref{f2.9}), then proof of divergence of
the expansion~(\ref{f1.3}), (\ref{f1.4}) may be done
analogously to one-dimensional case. But now some
complications will arise because of many-component character
of the variables $\vec x_i$.

So as earlier we will consider the contributions into $a_n$
proportional to $g^l$, denoting them as $a_n^l$. We treat at
the beginning $g$ as independent variable, so there are no
cancellations between $a_n^l$ with different $l$. To show
divergence of the series~(\ref{f1.4}) it is enough to show
divergence only for any special values of $\vec q\,',\; \vec q$,
e.g., for $\vec q\,' = \vec q$.

Let us evaluate contribution into $|a_n^l(\vec q, \vec q)|$ of
the form
  \begin{eqnarray} \label{f3.4}
  &&\frac{1}{2^{n-l}} \left| \int \limits_0^1 {\cal D}_n \eta\;
  \Delta_n \Delta_{n-1} \dots \Delta_{l+1}
  V(\vec x_l) \dots V(\vec x_1) \right| =
  \nonumber \\ &&
  \frac{1}{2^{n-l}} \left| \int \limits_0^1 {\cal D}_n \eta\;
  \sum_{\nu_n=1}^3 \dots \sum_{\nu_{l+1}=1}^3
  \frac{\partial^{2(n-l)}}{\partial x_{n, \nu_n}^2 \dots
                           \partial x_{l+1, \nu_{l+1}}^2}
  V(\vec x_l) \dots V(\vec x_1) \right|_{\vec q\,' =\vec q}.
  \end{eqnarray}
Due to transition to the derivatives of type
  $$V^{(m_1,m_2,m_3)}(\vec x_j) \equiv \frac{\partial^{m_1+m_2+m_3}
       V(\vec x_j)}{\partial x^{m_1}_{j_1} \partial x^{m_2}_{j_2}
       \partial x^{m_3}_{j_3}} $$
$\eta$-factors of the form $\eta_{n-1}^{k_{n-1}} \dots
\eta_{1}^{k_{1}}$ arise, which after integration in case
$\vec q\,' =\vec q$ give additional factor varying from
$n!(n-1)!/(2n-1)!$ to 1. This factor is not essential for
asymptotic estimates, because it does not affect factorial
growth. We take minimal possible value of this factor and get
estimate from below
  \begin{equation} \label{f3.5}
  \frac{1}{2^{n-l}} \frac{(n-1)!}{(2n-1)!} \left|
  \sum_{k_1=0}^{n-l} \sum_{k_2=0}^{n-l-k_1}
  \frac{(n-l)!}{k_1! k_2! k_3!}
  \frac{\partial^{2(n-l)} V^l(q)}
  {\partial q_1^{2k_1} \partial q_2^{2k_2} \partial q_3^{2k_3}}
  \right|_{k_3=n-l-k_1-k_2}.
  \end{equation}
For the sake of simplicity  we take among all terms of~(\ref{f3.5})
only that ones, in which differentiating with respect to only
one component of the vector $\vec q$, e.g., $q_1$, presents. Then
in~(\ref{f3.5}) one should put $k_1=n-l, \; k_2=k_3=0$ and
produce differentiation. One gets after this
  \begin{equation} \label{f3.6}
  \frac{1}{2^{n-l}} \frac{(n-1)!}{(2n-1)!} \left|
  \sum_{\{m_i\}} \frac{(2(n-l))!}{m_1! \dots m_l!}
  V^{(m_1,0,0)}(q) \dots V^{(m_l,0,0)}(q) \right|.
  \end{equation}
Notation $\sum_{\{m_i\}}$ coincides with one in~(\ref{f2.14}).

Taking into account the Taylor expansion (its validness at any
domain is assumed)
  \begin{eqnarray} \label{f3.7}
  V(\vec x)&=&
  \sum_{m_1=0}^{\infty} \sum_{m_2=0}^{\infty} \sum_{m_3=0}^{\infty}
  \frac{V^{(m_1,m_2,m_3)}(\vec q)}{m_1!m_2!m_3!}
  (x_1-q_1)^{m_1} (x_2-q_2)^{m_2} (x_3-q_3)^{m_3} =
  \nonumber \\ &&
  \sum_{m_1=0}^{\infty} \sum_{m_2=0}^{\infty} \sum_{m_3=0}^{\infty}
  C_{m_1,m_2,m_3}(\vec q)
  (x_1-q_1)^{m_1} (x_2-q_2)^{m_2} (x_3-q_3)^{m_3}
  \end{eqnarray}
one can rewrite~(\ref{f3.6}) in the form
  \begin{equation} \label{f3.8}
  \frac{1}{2^{n-l}} \frac{(n-1)!(2(n-l))!}{(2n-1)!} \left|
  \sum_{\{ m_i \}} C_{m_1,0,0}(\vec q) \dots C_{m_l,0,0}(\vec q) \right|.
  \end{equation}
From terms of the sum $\sum_{\{ m_i \}}$ we choose only one, in
which all $m_i$ are equal to $[2(n-l)/l]$ and consider
$l=[n\alpha]$ with $0 < \alpha <1/2$. Then we get
  \begin{equation} \label{f3.9}
  |a_n^{[n\alpha]}(\vec q,\vec q)| \sim
  \frac{(n-1)![2n(1-\alpha)]!}{2^{n-[n\alpha]}(2n-1)!}
  \left( C_{[2(1-\alpha)/\alpha],0,0}(\vec q) \right)^{[n\alpha]}
  \sim \Gamma(n(1-2\alpha)).
  \end{equation}

\subsection{Proof of divergence}

One is to show now that there are no essential cancellations
between different contributions into $a_n^l$ which diminish
factorial contributions of type~(\ref{f3.9}). Such cancellations
cannot be caused by different signs of derivatives
$V^{(m_1,m_2,m_3)}$ of different orders $m_\nu$. Reasonings
proving this fact almost exactly repeat reasonings of
previous Section. Only notations become slightly
complicated. Instead of~(\ref{f2.18}) one should write
  \begin{equation} \label{f3.10}
  |a_n^l(\vec q,\vec q)|= \left|
  \sum_{\{ m_{i,1} \}} \sum_{\{ m_{i,2} \}} \sum_{\{ m_{i,3} \}}
  A^{nl}_{\{m_{i,\nu} \}}  V^{(m_{11},m_{12},m_{13})}(\vec q)
                     \dots V^{(m_{l1},m_{l2},m_{l3})}(\vec q) \right|,
  \end{equation}
and in expression analogous to~(\ref{f2.20}) instead of expansion
in $q^k$  one should take expansion in the structures
$q_1^{k_1} q_2^{k_2} q_3^{k_3}$ and consider the coefficients
for every set $\{ k_{\nu} \}$.

In the three-dimensional case one more cause of cancellations
of contributions arises. It is following. If the potential
is harmonic function, i.e.,
  \begin{equation} \label{f3.11}
  \Delta V =0,
  \end{equation}
then, as it is clear from~(\ref{f3.3}), many of addends in $a_n$
are equal to zero. One is to show that it is enough of the
contributions of the terms remaining different from zero
to keep the asymptotic behavior~(\ref{f3.9}).

In~(\ref{f3.6}) the coefficient
  $$\frac{(2(n-l))!}{m_1! \dots m_l!} $$
has meaning of number of terms of the form $V^{(m_1,0,0)}(\vec q)
\dots V^{(m_l,0,0)}(\vec q)$ in $\sum_{\{ m_{i} \}}$. We chose
$m_i=[2(n-l)/l]$ for every $i$. Let us evaluate how many
terms among total number
  $$\frac{(2(n-l))!}{([2(n-l)/l]!)^l} $$
are different from zero, if $V(\vec q)$ satisfies
Eq.~(\ref{f3.11}). Here $l$ multipliers $V(\vec q)$ are
under action of $n-l$ operators $\Delta_i$ (we trace only
differentiating with respect to $x_{i1}$). The contribution will
be different from zero if every multiplier $V(\vec x)$ in it
is under action of differential operator which includes not more
then one derivative from every operator $\Delta_{l+1}, \dots,
\Delta_n$ (see~(\ref{f3.4})), i.e.,
  \begin{equation} \label{f3.12}
  \frac{\partial^{[2(n-l)/l]}}{\partial x_{i_1,1} \dots
                           \partial x_{i_{[2(n-l)/l]},1}},
  \end{equation}
where $i_j=l+1, \dots, n$ and there are no equal numbers among
$i_j$. Let such differential operator acts on $V(\vec x_1)$.
The set $\{ i_j \}$ can be chosen by
  $$\frac{(n-l)!}{(n-l-m_i)! m_i!}$$
ways. Considering in such manner $l/2$ multipliers $V(\vec x_i)$
(we mean $l$ as even for simplicity and $m_i=[2(n-l)/l]$ here)
we get the factor
  \begin{equation} \label{f3.13}
  \frac{(n-l)!}{(n-l-m_i)! m_i!} \frac{(n-l-m_i)!}{(n-l-2m_i)! m_i!}
  \dots \frac{(n-l-(\frac{l}{2}-1)m_i)!}{(n-l-\frac{l}{2}m_i)! m_i!} =
  \frac{(n-l)!}{(m_i!)^{l/2}}.
  \end{equation}
Next $l/2$ multipliers $V(\vec x_i)$ are under action of derivatives
from all operators $\Delta_{l+1}, \dots,$ $\Delta_n$, which was not used
before. This gives one more factor~(\ref{f3.13}). As a result we obtain
the estimate of number of terms different from zero
  \begin{equation} \label{f3.14}
  \frac{((n-l)!)^2}{(m_i!)^l} = \frac{((n-l)!)^2}{([2(n-l)/l]!)^l}
  \end{equation}
(not all contributions are taken into account here, but for our
purposes this underestimate is quite sufficient). For $l=[n\alpha]$
we get from~(\ref{f3.14})
  \begin{equation} \label{f3.15}
  \frac{([n(1-\alpha)]!)^2}{([2(1-\alpha)/\alpha]!)^{[n\alpha]}}
  \sim [2n(1-\alpha)]!.
  \end{equation}
I.e., taken into consideration different from zero contributions
in~(\ref{f3.14}) really provide asymptotic behavior~(\ref{f3.9})
causing divergence of the expansion~(\ref{f1.4}).

Eq.~(\ref{f3.15}) takes into account only part of increasing
contributions. Maximal growth corresponds to the case
$\alpha \to 0$. Assuming slight dependence of $\alpha$ on $n$ and
putting $\alpha = 1/\log n$ one gets, so as in previous
Section, that for $n \to \infty$
  \begin{equation} \label{f3.16}
  |a_n| \sim n!.
  \end{equation}
Such growth takes place for every potential excluding polynomial
ones. If $V(\vec q)$ is polynomial of order $L$, then
$C_{m_1,m_2,m_3}=0$ for $\sum\limits_{\nu=1}^3 m_{\nu} > L$.
So, there is boundary from below for possible values of $l$:
$l \ge 2n/(L+2)$. Maximal growth of $a_n^l$ takes place when
$l_m=[2n/(L+2)]$. This leads to estimate
  \begin{equation} \label{f3.17}
  |a_n| \sim \Gamma \left( n\frac{L-2}{L+2} \right),
  \end{equation}
which coincides with corresponding result for one-dimensional
theory. For $L \le 2$ (harmonic oscillator, linear potential,
free case) the expansion converges. For $L \ge 3$ (anharmonic
oscillator) it is divergent.

So as in one-dimensional case, for the potentials determined
by analytic, but not entire functions, i.e., by functions for
which Taylor series~(\ref{f3.7}) have finite convergence range,
one can prove divergence by rather simple way. Consider
  \begin{eqnarray} \label{f3.18}
  a_n^1(\vec q,\vec q) &=&
  - \frac{1}{2^{n-1}} \int \limits_0^1 {\cal D}_n \eta\;
  \Delta_n \Delta_{n-1} \dots \Delta_2 V(\vec x_1) =
  \nonumber \\ &&
  -\frac{1}{2^{n-1}} \int \limits_0^1 {\cal D}_n \eta\;
  \eta_{n-1}^2 \eta_{n-2}^4 \dots \eta_1^{2(n-1)}
  \times \nonumber \\ &&
  \sum_{k_1=0}^{n-1} \sum_{k_2=0}^{n-1-k_1}
  \frac{(n-1)!}{k_1! k_2! k_3!} V^{(2k_1,2k_2,2k_3)}(\vec q)
  \Biggl. \Biggr|_{k_3=n-1-k_1-k_2}.
  \end{eqnarray}
If $\sum\limits_{\nu=1}^3 2k_{\nu} = 2(n-1) \to \infty$
then the coefficients of the Taylor series for $V(\vec q)$
behave themselves as
  \begin{equation} \label{f3.19}
   V^{(2k_1,2k_2,2k_3)}(\vec q) \sim
  \frac{(2k_1)! (2k_2)! (2k_3)!}
  {R_1^{2k_1}(\vec q) R_2^{2k_2}(\vec q) R_3^{2k_3}(\vec q)},
  \end{equation}
where $R_{\nu}(\vec q)$ are conjugated convergence ranges of the
expansion~(\ref{f3.7}) (here we mean that for all $\nu$:
$0 < R_{\nu}(\vec q) < \infty$). This gives estimate for
$n \to \infty$
  \begin{equation} \label{f3.20}
  |a_n^1(\vec q,\vec q)| \sim \frac{(n-1)!}{2^{n-1} (2n-1)}
  \frac{1}{R(\vec q)^{2(n-1)}} \sim n!.
  \end{equation}
Here $R(\vec q) = \max \{ R_1(\vec q), R_2(\vec q),  R_3(\vec q) \}$.

If $V(\vec q)$ is harmonic function then $a_n^1(\vec q, \vec q) =0$.
Nevertheless, in this case one can easily calculate
$a_n^2(\vec q, \vec q)$
  \begin{eqnarray} \label{f3.21}
  &&a_n^2(\vec q,\vec q) =
  \frac{1}{2^{n-2}} \int \limits_0^1 {\cal D}_n \eta\;
  \sum_{\nu_n=1}^3 \dots \sum_{\nu_3=1}^3
  \frac{\partial^{n-2}V(\vec x_2)}{\partial x_{n, \nu_n} \dots
                           \partial x_{3, \nu_3}}
  \frac{\partial^{n-2}V(\vec x_1)}{\partial x_{n, \nu_n} \dots
                           \partial x_{3, \nu_3}}
  \Biggl. \Biggr|_{\vec x_i=\vec q} =
  \nonumber \\ &&
  \frac{1}{2^{n-2}} \int \limits_0^1 {\cal D}_n \eta\;
  \eta_{n-1}^2 \eta_{n-2}^4 \dots
  \eta_3^{2(n-3)} \eta_2^{2(n-2)} \eta_1^{n-2}
  \times \nonumber \\ &&
  \sum_{k_1=0}^{n-2} \sum_{k_2=0}^{n-2-k_1}
  \frac{(n-2)!}{k_1! k_2! k_3!}
  \left( V^{(k_1,k_2,k_3)}(\vec q) \right)^2
  \Biggl. \Biggr|_{k_3=n-2-k_1-k_2} =
  \nonumber \\ &&
  \frac{1}{2^{n-2}} \frac{(n-2)!}{(2(n-1))!}
  \sum_{k_1=0}^{n-2} \sum_{k_2=0}^{n-2-k_1}
  \frac{(n-2)!}{k_1! k_2! k_3!}
  \left( V^{(k_1,k_2,k_3)}(\vec q) \right)^2
  \Biggl. \Biggr|_{k_3=n-2-k_1-k_2}.
  \end{eqnarray}
Putting $k_1=n-2, \; k_2=k_3=0$ we obtain estimate for $n \to \infty$
  \begin{equation} \label{f3.22}
  |a_n^2(\vec q,\vec q)| > \frac{(n-2)!}{2^{n-2} (2(n-1))!}
  \left( V^{(n-2,0,0)}(\vec q) \right)^2 \sim
  \frac{(n-2)!}{2^{n-2} (2(n-1))!}
  \left( \frac{(n-2)!}{R_1^{n-2}(\vec q)} \right)^2 \sim n!.
  \end{equation}
Hence, for harmonic potentials factorial growth of the
coefficients $a_n$ takes place too.

So, analogously to the one-dimensional case, in the three-dimensional
space the Schwinger --- DeWitt expansion is divergent for all
potentials (excluding trivial polynomials of order not higher
then two) if the charge $g$ is considered as independent variable.
If the charge is considered as fixed parameter, then for some
kinds of the potentials and for some discrete values of the
charge the expansion~(\ref{f1.3})--(\ref{f1.4}) may be convergent.

\section{Conclusion}

The results of our research may be summarized as following.

If we consider for the beginning the coupling constant $g$ of
continuous potential
$V(q)$ as independent variable, then the coefficients $a_n$
of representation~(\ref{f1.3})--(\ref{f1.4})
for the evolution operator kernel increase for $n \to \infty$ as
  $$a_n \sim \Gamma \left( n \frac{L-2}{L+2} \right)$$
for the potentials being expressed via the polynomial of order
$L$ and as
  $$a_n \sim n!$$
for other ones. Hence, from this viewpoint the Schwinger --- DeWitt
expansion is divergent for all potentials excluding polynomials
of order not higher then two.

This is not surprising fact. The expansion~(\ref{f1.3}), (\ref{f1.4})
is usually considered as asymptotic. What about our considerations
we can see that conclusion
about divergence of the series~(\ref{f1.4}) is valid only if
we treat the charge $g$ as independent variable. But if the
charge is treated as fixed parameter, then
proof of divergence becomes not valid because of possibility of
cancellations for terms with different powers of $g$ in
$a_n(q',q)$. So, there is opportunity to avoid divergence.
And really some potentials for which the expansion~(\ref{f1.4})
is convergent for some discrete values of the charge $g$ are exist.
Examples of such potentials are known from
previous papers~\cite{S3,TMF2}.
In this case the function $F$ is analytic function
of variable $t$ at $t=0$ contrary to the case of independent
charge, when the point $t=0$ is essential singular point of $F$.

Discreteness of the charge for
the class of the potentials for which the
expansion is convergent, probably, may be connected with
discreteness of the charge in the nature. In this correspondence,
the potentials of this class are of special interest.
Operating with them we get rid of some kind of divergences
in the theory and, at the same time, have a theory with
discrete charge. So, it seems to be necessary to look for other
potentials of this class and study them carefully.


\begin{thebibliography} {99}

\bibitem{Sch}
J.~Schwinger, {\it Phys.~Rev.} {\bf 82} (1951) 664.

\bibitem{DeW1}
B.~S.~DeWitt, {\it Dynamical Theory of Groups and Fields} (Gordon \& Breach,
New York, 1965).

\bibitem{DeW2}
B.~S.~DeWitt, {\it Phys.~Rep.} {\bf 19} (1975) 297.

\bibitem{BOG}
A.~O.~Barvinsky, T.~A.~Osborn and Yu.~V.~Gusev, {\it J.~Math.~Phys.} {\bf 36}
(1995) 30.

\bibitem{OF}
T.~A.~Osborn and Y.~Fujiwara, {\it J.~Math.~Phys.} {\bf 24} (1983) 1093.

\bibitem{BW}
C.~M.~Bender and T.~T.~Wu, {\it Phys.~Rev.} {\bf 184} (1969) 1231.

\bibitem{Lip}
L.~N.~Lipatov, {\it Zh.E.T.F.} {\bf 72} (1977) 411.

\bibitem{PSS}
V.~S.~Popov, A.~V.~Sergeev, and A.~V.~Shcheblykin, {\it Zh.E.T.F.}
{\bf 102} (1992) 1453.

\bibitem{S1}
V.~A.~Slobodenyuk, {\it Z.~Phys.} {\bf C 58} (1993) 575.

\bibitem{S2}
V.~A.~Slobodenyuk, {\it Theor. Math. Phys.} {\bf 105} (1995) 1387,
[hep-th/9412001].

\bibitem{Martin}
A.~Martin, {\it Phys.~Rep.} {\bf 134} (1986) 305.

\bibitem{S3}
V.~A.~Slobodenyuk, Preprint IHEP 95--70, Protvino, 1995, [hep-th/9506134].

\bibitem{TMF2}
V.~A.~Slobodenyuk, {\it Theor. Math. Phys.} {\bf 109} (1996) 1302.

\bibitem{MPLA2}
V.~A.~Slobodenyuk, {\it Mod. Phys. Lett.} {\bf A 11} (1996) 1729,
[hep-th/9605188].

\end{thebibliography}
\end{document}